\documentstyle[12pt,twoside]{article} 

\def\ad{a^{\dag}}
\def\hO{\hat O}\def\h1{\hat 1}

\def\ord{\mbox{\bf :}}
\def\hr{\hat\varrho}
\def\<{\langle}\def\>{\rangle}
\def\cj#1{\overline{#1}}

\def\<{\langle}\def\>{\rangle}
\def\ev#1{\< #1\>}
\def\pni{\par\noindent}
\tolerance = 10000
\title{Trace forms for the generalized Wigner functions}
\author{G. M. D'Ariano\protect$^{a,b}$ and M. F. Sacchi\protect$^{a}$\\
\\
\protect$^a$ {\em Dipartimento di Fisica \lq Alessandro Volta\rq}, \\
{\em Universit\`a degli Studi di Pavia},\\ {\em 
via A. Bassi 6, I-27100 Pavia, Italy}\\
\\
\protect$^b$ {\em Department of Electrical and Computer Engineering},\\
{\em Department of Physics and Astronomy},\\
{\em Northwestern University, Evanston, IL 60208, USA}}
\date{Ricevuto: }
\begin{document}
\maketitle\thanks{1996 PACS numbers: 03.65.Bz, 42.50.Dv}
\begin{abstract}
We derive simple formulas connecting the generalized Wigner functions 
for $s$-ordering with the density matrix, and {\em vice-versa}.
These formulas proved very useful for quantum mechanical applications, 
as, for example, for connecting master equations with Fokker-Planck
equations, or for evaluating the quantum state from Monte Carlo 
simulations of Fokker-Planck equations, and finally for 
studying positivity of the generalized Wigner functions in the complex
plane.
\medskip
\par\noindent{\bf RIASSUNTO}:
In questo lavoro deriviamo semplici formule che connettono
direttamente le funzioni generalizzate di Wigner con la rispettiva
matrice densit\`a. Queste formule sono molto utili in applicazioni
quantomeccaniche, come, ad esempio, nel connettere master equations
con equazioni di Fokker-Plank, o per determinare lo stato 
quantistico da simulazioni di equazioni di Fokker-Plank, o, infine, 
per determinare la positivit\`a delle funzioni di Wigner generalizzate
sul piano complesso.
\end{abstract}
\vfill\eject
Since the Wigner's pioneering work \cite{wig}, generalized phase-space 
techniques have proved very useful in various branches of physics \cite{phs}. 
As a method for expressing the density operator in terms of c-number
functions, the Wigner functions often lead to considerable simplification 
of the quantum equations of motion, as for example, transforming operator 
master equations into more amenable Fokker-Planck differential equations 
(see, for example, Ref. \cite{gard}). 
By the Wigner function one can express quantum-mechanical expectation 
values in form of averages over the complex plane (the classical
phase-space), the Wigner function playing the role of a 
c-number quasi-probability distribution, which generally can also
have negative values.
More precisely, the original Wigner function allows to easily evaluate
expectations of symmetrically ordered products of the field operators, 
corresponding to the Weyl's quantization procedure
\cite{weyl}. However, with a slight change of the
original definition, one defines generalized 
$s$-ordered Wigner function $W_s(\alpha,\cj{\alpha})$, as follows \cite{cgl2}
\begin{eqnarray}
W_s(\alpha,\cj{\alpha})\doteq\int\frac{d^2\lambda}{\pi^2}
e^{\alpha\cj{\lambda}-\cj{\alpha}\lambda+{s\over2}|\lambda|^2}
\mbox{Tr}[D(\lambda)\hr]\;\label{Ws}\;
\end{eqnarray}
where the integration is performed on the complex plane with
measure $d^2\lambda=d{\hbox {Re}}\lambda\,d{\hbox {Im}}\lambda$,
$D(\alpha )=e^{\alpha \ad -\cj{\alpha }a}$ denotes the
displacement operator, and $a$ and $a^{\dag}$ ($[a,a^{\dag}]=1$) are the
annihilation and creation operators of the field mode of interest.
Then, using the Wigner function in Eq. (\ref{Ws}) one can evaluate 
$s$-ordered expectation values of the field operators 
through the following relation 
\begin{eqnarray}
\mbox{Tr}[\ord (a^{\dag})^na^m\ord_s \hr]
=\int d^2\alpha\,W_s(\alpha,
\cj{\alpha})\,\cj{\alpha}^n\alpha^m\;.\label{ex}
\end{eqnarray}
It is easy to show that the particular cases 
$s=-1,0,1$ lead to {\em antinormal}, {\em symmetrical}, and
{\em normal} ordering, respectively, in which cases the generalized
Wigner function $W_s(\alpha, \cj{\alpha})$ historically was denoted
with the following symbols and names
\begin{eqnarray}
W_s(\alpha,\cj{\alpha})=
\cases{
{1\over\pi}Q(\alpha,\cj{\alpha})\quad&\mbox{ for }s=-1
\mbox{ ``$Q$-function''}\cr
W(\alpha,\cj{\alpha})\quad&\mbox{ for }s=0\mbox{ (usual Wigner function)}
\cr P(\alpha,\cj{\alpha})\quad&\mbox{ for }s=1\mbox{ ``$P$-function''}
\cr}\label{3}\end{eqnarray}
For the normal ($s=1$) and antinormal ($s=-1$) orderings, the
following two simple relations between the generalized Wigner 
function and the density matrix are well known 
\begin{eqnarray}
&&Q(\alpha,\cj{\alpha})\equiv\<\alpha|\hr|\alpha\>\;,\label{QQ}\\
&&\hr=\int d^2\alpha\,P(\alpha,\cj{\alpha})\,|\alpha\>\<\alpha|\;,
\label{P}\end{eqnarray}
where $|\alpha \>$ denotes the customary coherent state 
$|\alpha \>=D(\alpha )|0\>$,  $|0\>$ being the vacuum state of the field.
Among the three particular representations (\ref{3}), it is also well
known that the Q-function is positively definite and infinitely
differentiable (it actually represents the probability distribution
for ideal joint measurements of position and momentum of the
harmonic oscillator: see, for example, Ref. \cite{darank}).
On the other hand, the P-function is known to be possibly highly singular, 
and the only pure states for which it is positive are the coherent 
states \cite{cah}. Finally, the usual Wigner function
has the remarkable property of providing the probability distributions 
of the quadratures of the field in form of marginal distributions, namely
\begin{eqnarray}
\int d{\hbox {Im}}\alpha\,W(\alpha e^{i\phi},\cj{\alpha }e^{-i\phi})
={}_{\phi}\ev{{\hbox {Re}}\alpha|\hr|{\hbox {Re}}\alpha}_{\phi}\;,
\end{eqnarray}
where $|x \>_{\phi}$ stands for the eigenstates of the field 
quadrature $\hat X_{\phi}=(\ad e^{i\phi}+\hbox{h.c.})/2$
(any couple of conjugated quadratures $\hat X_{\phi}$, $\hat X_{\phi+\pi/2}$,
with $[\hat X_{\phi},\hat X_{\phi+\pi/2}]=i/2$, are equivalent to the 
position and momentum of a harmonic oscillator). Usually, 
negative values of the Wigner function are viewed as signature of a 
nonclassical state (one of the more eloquent examples is given by 
the Schr\"odinger-cat states \cite{cat} 
whose Wigner function is characterized by rapid oscillations around the 
origin of the complex plane). 
From Eq. (\ref{Ws}) one can see that all $s$-ordered Wigner functions
are related to each other through the convolution relation
\begin{eqnarray}
W_s(\alpha,\cj{\alpha})&=&\int d^2\beta\,
W_{s'}(\beta,\cj{\beta})
\frac{2}{\pi(s'-s)}\exp\left(-\frac{2}{s'-s}|\alpha-\beta|^2\right)\label{conv}
\\
&=&\exp\left(\frac{s'-s}{2}
\frac{\partial^2}{\partial\alpha\partial\cj{\alpha}}\right)
W_{s'}(\alpha,\cj{\alpha})\;,\quad (s'>s)\;.
\end{eqnarray}
Eq. (\ref{conv}) shows the positiveness of the generalized Wigner function
for $s<-1$, as a consequence of the positiveness of the Q-function. 
From a qualitative point of view, the maximum value 
of $s$ keeping the generalized Wigner functions as positive can be
considered as an indication of the classical nature of the physical state.
\par In this paper we present three equivalent trace forms that connect 
$s$-ordered Wigner functions with the density matrix. 
They are the following:
\begin{eqnarray}
W_s(\alpha,\cj{\alpha})=\frac{2}{\pi(1-s)}
e^{-\frac{2}{1-s}|\alpha|^2}\mbox{Tr}\left[\left(\frac{s+1}{s-1}
\right)^{a^{\dag}a}e^{\frac{2}{1-s}\cj{\alpha}a}\hr
e^{\frac{2}{1-s}\alpha a^{\dag}}\right]\;,\label{W1}\\
=\frac{2}{\pi(1-s)}
e^{\frac{2}{1+s}|\alpha|^2}\mbox{Tr}\left[\left(\frac{s+1}{s-1}
\right)^{a^{\dag}a}e^{-\frac{2}{1+s}\alpha a^{\dag}}\hr
e^{-\frac{2}{1+s}\cj{\alpha}a}\right]\;,\label{W2}\\
=\frac{2}{\pi(1-s)}e^{
-\frac{2s}{1-s^2}|\alpha|^2}\mbox{Tr}
\left[\left(\frac{s+1}{1-s}\right)^{{1\over2}a^{\dag}a}\!\!\!\!\!
D\left(\frac{2\alpha}{\sqrt{1-s^2}}\right)
\left(\frac{s+1}{1-s}\right)^{{1\over2}a^{\dag}a}\!\!\!\!\! 
(-)^{a^{\dag}a}
\hr\right]\;.\label{W3}
\end{eqnarray}
Eqs.(\ref{W1}-\ref{W2}) can be compared with the Cahill-Glauber 
formula \cite{cgl2}
\begin{eqnarray}
W_s(\alpha,\cj{\alpha})=\frac{2}{\pi(1-s)} 
\mbox{Tr}\left\{\ord \exp \left[ - 
\frac{2}{1-s}(\cj\alpha- a^{\dag}) (\alpha- a) 
\right] \ord \hr\right\}\;,\label{not}
\end{eqnarray}
where the colons denote the usual normal ordering; Eq. (\ref{W3}) represents a 
generalization of the formula \cite{cgl}
\begin{eqnarray}
W(\alpha,\cj{\alpha})=\frac{2}{\pi}\mbox{Tr}
\left[\hr D(2\alpha )\exp(i\pi a^{\dag}a)\right]\;.
\end{eqnarray}
\par Vice versa, the density matrix can be recovered from 
the generalized Wigner functions using the following expression
\begin{eqnarray}
\hr\!=\!\frac{2}{1+s}\!\int\! d^2\alpha W_s(\alpha,\cj{\alpha})
e^{-\frac{2}{1+s}|\alpha|^2}\exp\left(
\frac{2\alpha}{1+s} a^{\dag}\right)
\!\!\left(\frac{s-1}{s+1}\right)^{a^{\dag}a}\!\!\!\!
\exp\left(\frac{2\cj{\alpha}}{1+s}a\right).\!\label{mybest2}
\end{eqnarray}
The proof of our statements requires the following identity
\begin{eqnarray}
e^{a^{\dag}\partial_{\cj{\alpha}}}|0\>\< 0|
e^{a\partial_{\alpha}}\Bigg\vert_{\alpha=\cj{\alpha}=0}
e^{|\alpha|^2+\cj{\alpha}\lambda-\alpha\cj{\lambda}-{1\over2}
|\lambda|^2}=D(\lambda)\;,\label{disp}
\end{eqnarray}
which is proved in the Appendix. Then, through the following steps:
\begin{eqnarray*}
&&W_s(\alpha,\cj{\alpha})=\int\frac{d^2\lambda}{\pi^2}
e^{\alpha\cj{\lambda}-\cj{\alpha}\lambda+{s\over2}|\lambda|^2}
\mbox{Tr}[D(\lambda)\hr]\\
&&=\int\frac{d^2\lambda}{\pi^2}e^{\alpha\cj{\lambda}-
\cj{\alpha}\lambda}\mbox{Tr}\left[e^{a^{\dag}\partial_{\cj{\beta}}}
|0\>\< 0|e^{a\partial_{\beta}}\hr\right]\Bigg\vert_{\beta=\cj{\beta}
=0}e^{|\beta|^2+\cj{\beta}\lambda-\beta\cj{\lambda}+{1\over2}(s-1)
|\lambda|^2}\\ 
&&={2\over{\pi(1-s)}}\mbox{Tr}
\left[e^{a^{\dag}\partial_{\cj{\beta}}}|0\>\< 0|e^{a\partial_{\beta}}
\hr\right]\Bigg\vert_{\beta=\cj{\beta}=0}
e^{-\frac{1+s}{1-s}|\beta|^2-\frac{2}{1-s}(|\alpha|^2-\alpha
\cj{\beta}-\cj{\alpha}\beta)}\\ &&
={2\over{\pi(1-s)}}e^{-\frac{2}{1-s}|\alpha|^2}\mbox{Tr}
\left[e^{\frac{2}{1-s}\alpha a^{\dag}}
\left(-\frac{1+s}{1-s}\right)^{a^{\dag}a}e^{\frac{2}{1-s}
\cj{\alpha}a}\hr\right]\;,
\end{eqnarray*}
one proves Eq. (\ref{W1}). Continuing from the last result we have
\begin{eqnarray*}
&&\ \ W_s(\alpha,\cj{\alpha})={2\over{\pi(1-s)}}
e^{-\frac{2}{1-s}|\alpha|^2}\times \\
&&\!\!\!\!\!\!\!\!\!\!\!\!\!\!\!\mbox{Tr}
\!\left[\left(\frac{1+s}{1-s}\right)^{{1\over2}a^{\dag}a}
\!\!\left(\frac{1-s}{1+s}\right)^{{1\over2}a^{\dag}a}
\!e^{\frac{2}{1-s}\alpha a^{\dag}}
\!\left(-\frac{1+s}{1-s}\right)^{a^{\dag}a}
\!e^{\frac{2}{1-s}\cj{\alpha}a}
\!\left(\frac{1-s}{1+s}\right)^{{1\over2}a^{\dag}a}
\!\!\left(\frac{1+s}{1-s}\right)^{{1\over2}a^{\dag}a}\!\!\hr\right]\\
&&\!\!\!\!\!\!\!\!\!\!\!\!\!\!\!={2\over{\pi(1-s)}}e^{-\frac{2}{1-s}|\alpha|^2}
\mbox{Tr}\left[\left(\frac{1+s}{1-s}\right)^{{1\over2}a^{\dag}a}
e^{\frac{2}{\sqrt{1-s^2}}\alpha a^{\dag}}(-)^{a^{\dag}a}
e^{\frac{2}{\sqrt{1-s^2}}\cj{\alpha}a}
\left(\frac{1+s}{1-s}\right)^{{1\over2}a^{\dag}a}\!\!\hr\right]\\
&&\!\!\!\!\!\!\!\!\!\!\!\!\!\!\!={2\over{\pi(1-s)}}e^{-\frac{2s}{1-s^2}
|\alpha|^2}\mbox{Tr}\left[\left(\frac{1+s}{1-s}\right)^{{1\over2}a^{\dag}a}
\!\!D\left(\frac{2\alpha}{\sqrt{1-s^2}}\right)
\left(\frac{1+s}{1-s}\right)^{{1\over2}a^{\dag}a}(-)^{a^{\dag}a}
\hr\right]\;,
\end{eqnarray*}
which proves Eq. (\ref{W3}). 
Eq. (\ref{W2}) is derived using the following identities
\begin{eqnarray*}
&&e^{\frac{2\alpha}{1-s}a^{\dag}}\left(\frac{s+1}{s-1}\right)^{
a^{\dag}a}e^{\frac{2\cj{\alpha}}{1-s}a}=\left(\frac{s+1}{s-1}
\right)^{a^{\dag}a}e^{-\frac{2\alpha}{1+s}a^{\dag}}e^{\frac{2\cj{
\alpha}}{1-s}a}\;,\\
&&e^{\frac{4|\alpha|^2}{1-s^2}}\left(\frac{s+1}{s-1}\right)^{
a^{\dag}a}e^{\frac{2\cj{\alpha}}{1-s}a}e^{-\frac{2\alpha}{1+s}
a^{\dag}}=e^{\frac{4|\alpha|^2}{1-s^2}}
e^{-\frac{2\cj{\alpha}}{1+s}a}\left(\frac{s+1}{s-1}\right)^{
a^{\dag}a}e^{-\frac{2\alpha}{1+s}a^{\dag}}\;.
\end{eqnarray*}
As a check, from Eqs. (\ref{W1}-\ref{W3}) one can easily recover the 
usual definition of the Wigner function (\ref{Ws}) for $s=0$, and
Eq. (\ref{QQ}) for the $Q$-function ($s=-1$), namely
\begin{eqnarray*}
W_{-1}(\alpha,\cj{\alpha})&=&{1\over\pi}e^{-|\alpha|^2}\mbox{Tr}
\left[(-O^+)^{a^{\dag}a}e^{\cj{\alpha}a}\hr e^{\alpha a^{\dag}}
\right]={1\over\pi}e^{-|\alpha|^2}\mbox{Tr}\left[|0\>\< 0|
e^{\cj{\alpha}a}\hr e^{\alpha a^{\dag}}\right]\nonumber\\
&=&{1\over\pi}Q(\alpha,\cj{\alpha})\;.
\end{eqnarray*}
The inversion formula (\ref{mybest2}) is obtained using Eq. (\ref{W3}) 
and the following formula \cite{cgl2} 
\begin{eqnarray}
\hO=\int\frac{d^2\alpha}{\pi}\mbox{Tr}[\hO D(\alpha)]
D^{\dag}(\alpha)\;,\label{tom}
\end{eqnarray}
that holds true for any Hilbert-Schmidt operator $\hO$, and hence 
for a (trace-class) density matrix. One has
\begin{eqnarray}
&&\left(\frac{s+1}{1-s}\right)^{{1\over2}a^{\dag}a}(-)^{a^{\dag}a}
\hr\left(\frac{s+1}{1-s}\right)^{{1\over2}a^{\dag}a}\nonumber\\
&=&
\int\frac{d^2\alpha}{\pi}\mbox{Tr}\left[D(\alpha)
\left(\frac{s+1}{1-s}\right)^{{1\over2}a^{\dag}a}(-)^{a^{\dag}a}
\hr\left(\frac{s+1}{1-s}\right)^{{1\over2}a^{\dag}a}
\right]D^{\dag}(\alpha)\nonumber\\
&=&\frac{4}{1-s^2}\int\frac{d^2\alpha}{\pi}W_s(\alpha,\cj{\alpha})
\frac{\pi(1-s)}{2}e^{\frac{2s}{1-s^2}{|\alpha|^2}}D^{\dag}
\left(\frac{2\alpha}
{\sqrt{1-s^2}}\right)\;.
\end{eqnarray}
Hence,
\begin{eqnarray*}
\hr\!=\!\frac{2}{1+s}\int\!d^2\alpha W_s(\alpha,\cj{\alpha})
e^{\frac{2s}{1-s^2}{|\alpha|^2}}\left(\frac{1-s}{1+s}\right)^{{1\over2}
a^{\dag}a}\!(-)^{a^{\dag}a}
D^{\dag}\left(\frac{2\alpha}{\sqrt{1-s^2}}\right)
\left(\frac{1-s}{1+s}\right)^{{1\over2}a^{\dag}a} \\
\!=\!\frac{2}{1+s}\int \!d^2\alpha W_s(\alpha,\cj{\alpha})
e^{-\frac{2}{1+s}{|\alpha|^2}}\left(\frac{1-s}{1+s}\right)^{{1\over2}a^{\dag}a}
e^{\frac{2\alpha}{\sqrt{1-s^2}}a^{\dag}}(-)^{a^{\dag}a}
e^{\frac{2\cj{\alpha}}{\sqrt{1-s^2}}a}
\left(\frac{1-s}{1+s}\right)^{{1\over2}a^{\dag}a}\!\!\! ,
\end{eqnarray*}
and then the result follows easily. In particular, for $s=0$ one has
the inverse of the Glauber formula
\begin{eqnarray}
\hr=2\int d^2\alpha  W(\alpha,\cj{\alpha}) D(2\alpha)
(-)^{a^{\dag}a}\;,
\end{eqnarray}
whereas for $s=1$ one recovers the relation (\ref{P}) that defines the
$P$-function.
\par The trace form of Eqs.(\ref{W1}-\ref{W2}-\ref{W3}) can be used
for an analysis of positivity of the Wigner function, usually a quite
difficult task, as confirmed in Ref. \cite{orlw}. In particular, from
Eq. (\ref{W1}) one can immediately see that for $s<1$ (namely, with
the only exception of the $P$-function) the $s$-Wigner function can become
negative, because the operator 
$e^{\frac{2}{1-s}\cj{\alpha}a}\hr e^{\frac{2}{1-s}\alpha a^{\dag}}$
is positive-definite, whereas the preceding factor $\left(\frac{s+1}{s-1}
\right)^{a^{\dag}a}$ is negative for $s<1$, and positivity is guaranteed 
only for products of positive operators. On the other hand, from Eq. 
(\ref{W3}) one can easily see that there is always a
state (the eigenstate of $a^{\dag}a$ with odd eigenvalue) that makes
the $s$-Wigner function at $\alpha=0$ negative for $s<0$.
\par The representations (\ref{W1},\ref{W2}) 
for the generalized Wigner functions also provide the easiest way to 
derive differential representations for boson operators acting on a density 
matrix. By defining, analogously to Eq. (\ref{Ws}), the generalized 
{\em Wigner symbol} for any operator $\hat O$,
\begin{eqnarray}
W_s(\alpha,\cj{\alpha}|\hat O)\doteq\int\frac{d^2\lambda}{\pi^2}
e^{\alpha\cj{\lambda}-\cj{\alpha}\lambda+{s\over2}|\lambda|^2}
\mbox{Tr}[D(\lambda)\hat O]
\;,
\end{eqnarray}
from Eqs. (\ref{W1},\ref{W2}) one immediately derives the relations
\begin{eqnarray}
&&W_s(\alpha,\cj{\alpha}|a\hr)=e^{-\frac{2}{1-s}|\alpha|^2}
\frac{1-s}{2}\,\partial_{\cj{\alpha}}\,e^{\frac{2}{1-s}|\alpha|^2}
W_s(\alpha,\cj{\alpha})\;,\nonumber\\&&=
\left(\alpha+\frac{1-s}{2}\,\partial_{\cj{\alpha}}\right)
W_s(\alpha,\cj{\alpha}) \\
&&W_s(\alpha,\cj{\alpha}|a^{\dag}\hr)=e^{\frac{2}{1+s}|\alpha|^2}
\left(-\frac{1+s}{2}\,\partial_{\alpha}\right)
e^{-\frac{2}{1+s}|\alpha|^2} W_s(\alpha,\cj{\alpha})\nonumber\\&&=
\left(\cj{\alpha}-\frac{1+s}{2}\,\partial_{\alpha}\right)
W_s(\alpha,\cj{\alpha})\;,
\end{eqnarray}
and analogous relations for right multiplication by the boson operator.
More generally, one can write a differential representation for any 
{\em super-operator}---i. e. right or left multiplication by an
operator $\hat O$---namely
\begin{eqnarray}
W_s(\alpha,\cj{\alpha}|\hat O\hr)\doteq F_s[\hat O\cdot] 
W_s(\alpha,\cj{\alpha})\;,
\qquad W_s(\alpha,\cj{\alpha}|\hr\hat O)\doteq F_s[\cdot\hat O]
W_s(\alpha,\cj{\alpha})\;,
\end{eqnarray}
where $\hat O\cdot$ and $\cdot\hat O$ denote left and right
multiplication by the operator $\hat O$, respectively, and 
$F_s$ are differential forms functions of $\alpha$, $\cj{\alpha}$,
$\partial_{\alpha}$ and $\partial_{\cj{\alpha}}$ with the following properties
\begin{eqnarray}
&&F_s[\hat O_1\hat O_2\cdot]=F_s[\hat O_1\cdot]F_s[\hat O_2\cdot]\;,
\label{mult1}\\
&&F_s[\cdot\hat O_1\hat O_2]=F_s[\cdot\hat O_2]F_s[\cdot\hat O_1
]\;.\label{mult2}\\
&&[F_s[\cdot\hat O_1],F_s[\hat O_2\cdot]]=0\;,\label{lrcomm}\\
&&F_s[\cdot\hat O]=\cj{F}_s[\hat O^{\dag}\cdot]\;.\label{hermitian}
\end{eqnarray}
The functional forms of the basic super-operators are summarized in 
Table~\ref{tab-differential}. 
The representations of $\cdot a^{\dag}$ and $\cdot a$ can be easily 
obtained from those of $a\cdot$ and $a^{\dag}\cdot$ using identities 
(\ref{hermitian}). Eq. (\ref{lrcomm}) is just the obvious 
statement that ``left multiplication commutes with right 
multiplication'' (for $a$ and $a^{\dag}$ this
corresponds to the identity $[\partial_{
\alpha}+\kappa\cj{\alpha},\partial_{\cj{\alpha}}+\kappa\alpha]=0$). 
\begin{table}[hbt]
\begin{center}
\begin{tabular}{|c|l|}
\hline\hline
Super-operator & $F_s$ \\ \hline\hline
$a\cdot$ & $\alpha+\frac{1-s}{2}\partial_{\cj{\alpha}}$\\ \hline
$a^{\dag}\cdot$ & $\cj{\alpha}-\frac{1+s}{2}\partial_{\alpha}$\\ \hline
$\cdot a$ & $\alpha-\frac{1+s}{2}\partial_{\cj{\alpha}}$\\ \hline
$\cdot a^{\dag}$ & $\cj{\alpha}+\frac{1-s}{2}\partial_{\alpha}$\\ \hline
$a\cdot a^{\dag}$ & $|\alpha|^2+\frac{1-s}{2}(1+\alpha\partial_{\alpha}+
\cj{\alpha}\partial_{\cj{\alpha}})+\left(\frac{1-s}{2}\right)^2
\partial_{\alpha\cj{\alpha}}$\\ \hline
$a^{\dag}\cdot a$ & $|\alpha|^2-\frac{1+s}{2}(1+\alpha\partial_{\alpha}+
\cj{\alpha}\partial_{\cj{\alpha}})+\left(\frac{1+s}{2}\right)^2
\partial_{\alpha\cj{\alpha}}$\\ \hline
$a^{\dag}a\cdot$ & $|\alpha|^2+{1\over2}[(1-s)\cj{\alpha}
\partial_{\cj{\alpha}}
-(1+s)\alpha\partial_{\alpha}-(1+s)-{1\over2}(1-s^2)
\partial_{\alpha\cj{\alpha}}$]\\ \hline
$\cdot a^{\dag}a$ & $|\alpha|^2+{1\over2}[(1-s)\alpha\partial_{\alpha}
-(1+s)\cj{\alpha}\partial_{\cj{\alpha}}-(1+s)-{1\over2}(1-s^2)
\partial_{\alpha\cj{\alpha}}]$\\ 
\hline\hline
\end{tabular}
\end{center}
\caption[fake]{Differential Wigner representation of some super-operators
\label{tab-differential}}\end{table}
Then, the differential representation of higher-order super-operators
is easily obtained from the composition rules (\ref{mult1}) and
(\ref{mult2}). Using the differential representation for Bose
super-operators, one can convert master equations into (possibly high
order) Fokker-Planck equations. For example, the master equation
of the damped harmonic oscillator (damping coefficient $\gamma$ and
thermal  photons $\bar n$)
\begin{eqnarray}
\partial_t\hr=-{\gamma \over 2}(\bar n +1)(\ad a\hr+\hr\ad a-2a\hr\ad)
-{\gamma \over 2}\bar n(a\ad \hr +\hr a\ad -2\ad \hr a)\;,
\end{eqnarray}
can be converted into the equivalent Fokker-Planck equation for the 
$s$-ordered Wigner function
\begin{eqnarray}
\partial_t W_s(\alpha,\cj{\alpha})={\gamma \over 2}
\left[\partial_{\alpha}\alpha +\partial_{\cj{\alpha}}\cj{\alpha }+
(2\bar n+1-s)
\partial_{\alpha \cj{\alpha}}\right]W_s(\alpha,\cj{\alpha})
\;.
\end{eqnarray}
For solving Fokker-Planck equations, one can use very efficient
Monte-Carlo Green-function simulation methods (see, for example,
Ref. \cite{our}), choosing the parameter $s$ such that both the 
Wigner function  and the diffusion coefficient remain positive during
the evolution.
Then, from the inversion Eq. (\ref{mybest2}) one can recover the
matrix elements $\< n|\hr|m\>$ of the operator $\hr$ in form of
Monte-Carlo integrals of Laguerre polynomials.
\par In conclusion, we have presented simple trace formulas that
connect the generalized Wigner functions with the density matrix, 
and {\em vice-versa}, and we have shown how they can be practically
used for: {\em i)} studying positivity of the generalized Wigner functions;
{\em ii)} connecting master equations with Fokker-Planck
equations; {\em iii)} evaluating the quantum state in Monte Carlo 
simulations of Fokker-Planck equations.
\section*{Appendix}
Proof of identity (\ref{disp}).
\pni From the relation
\begin{eqnarray}
\partial^n_{\cj{\alpha}}\partial^m_{\alpha}
\Bigg\vert_{\alpha=\cj{\alpha}=0}e^{|\alpha|^2}=\delta_{nm}n!\;,
\end{eqnarray}
one has
\begin{eqnarray}
&&e^{a^{\dag}\partial_{\cj{\alpha}}}|0\>\< 0|
e^{a\partial_{\alpha}}\Bigg\vert_{\alpha=\cj{\alpha}=0}
e^{|\alpha|^2}=\sum_{n,m=0}^{\infty}(a^{\dag})^n|0\>\< 0|
a^m\frac{\partial^n_{\cj{\alpha}}\partial^m_{\alpha}}{n!m!}
\Bigg\vert_{\alpha=\cj{\alpha}=0}e^{|\alpha|^2}\nonumber \\ &&=
\sum_{n=0}^{\infty}(a^{\dag})^n|0\>\< 0|a^n{1\over n!}=
\sum_{n=0}^{\infty}|n\>\< n|=\hat 1\;.
\end{eqnarray}
Hence, using the identities
\begin{eqnarray}
e^{a^{\dag}\partial_{\cj{\alpha}}}e^{\cj{\alpha}\lambda}=
e^{\lambda(a^{\dag}+\cj{\alpha})}e^{a^{\dag}
\partial_{\cj{\alpha}}}\;,\qquad
e^{a\partial_{\alpha}}e^{-\alpha\cj{\lambda}}=
e^{-\cj{\lambda}(a+\alpha)}e^{a\partial_{\alpha}}\;,
\end{eqnarray}
one obtains
\begin{eqnarray}
&&e^{a^{\dag}\partial_{\cj{\alpha}}}|0\>\< 0|
e^{a\partial_{\alpha}}\Bigg\vert_{\alpha=\cj{\alpha}=0}
e^{|\alpha|^2+\cj{\alpha}\lambda-\alpha\cj{\lambda}-{1\over2}
|\lambda|^2}\nonumber \\ &&=e^{-{1\over2}|\lambda|^2}
e^{\cj{\alpha}\lambda-\alpha\cj{\lambda}}e^{a^{\dag}\lambda}
e^{a^{\dag}\partial_{\cj{\alpha}}}|0\>\< 0|
e^{a\partial_{\alpha}}\Bigg\vert_{\alpha=\cj{\alpha}=0}
e^{|\alpha|^2}e^{-\cj{\lambda}a}\nonumber\\
&&=e^{-{1\over2}|\lambda|^2}
e^{a^{\dag}\lambda}e^{-a\cj{\lambda}}=D(\lambda)\;.
\end{eqnarray}
\section*{References}
\begin{description}
\bibitem[1]{wig} E. P. Wigner, Phys. Rev. {\bf 40}, 749 (1932).
\bibitem[2]{phs} {\em The Physics of Phase Space}, 
edited by Y. S. Kim and W. W. Zachary (Springer, Berlin, 1986).
\bibitem[3]{gard} C. W. Gardiner, {\em Quantum Noise} (Springer, 
Berlin, 1991).
\bibitem[4]{weyl} H. Weyl, {\em The Theory of Groups and Quantum Mechanics} 
(Dover, New York, 1950).
\bibitem[5]{cgl2} K. E. Cahill and R. J. Glauber, 
Phys. Rev. {\bf 177}, 1857 (1969).
\bibitem[6]{darank} For a tutorial review on concepts of quantum
measurements and applications to quantum optics, see: 
G. M. D'Ariano, {\em Quantum Estimation Theory
and Optical Detection}, in {\em Concepts and Advances in Quantum Optics and 
Spectroscopy of Solids}, ed. by T. Hakio\u{g}lu  and A. S. Shumovsky. 
(Kluwer Academic Publishers, Amsterdam 1996, in press).
\bibitem[7]{cah} K. E. Cahill, Phys. Rev. {\bf 138}, B1566 (1965).
\bibitem[8]{cat} G. M. D'Ariano, M. Fortunato and P. Tombesi, Nuovo 
Cimento B {\bf 110}, 1127 (1995).
\bibitem[9]{cgl} K. E. Cahill and R. J. Glauber, 
Phys. Rev. {\bf 177}, 1882 (1969).
\bibitem[10]{orlw} A. Or{\l}owski and A. W\"unsche, Phys. Rew. A, {\bf 48}, 
4697 (1993).
\bibitem[11]{our} G. M. D'Ariano, C. Macchiavello, and S. Moroni,
Mod. Phys. Lett. B {\bf 8}, 239 (1994).
\end{description}
\end{document}